# *WebScript* – A Scripting Language for the Web

Yin Zhang

Cornell Network Research Group (CNRG)

Department of Computer Science

Cornell University

Ithaca, NY 14853, U.S.A.

E-mail: `yzhang@cs.cornell.edu`

**Abstract**

*WebScript* is a scripting language for processing Web documents. Designed as an extension to Jacl, the Java implementation of Tcl, *WebScript* allows programmers to manipulate HTML in the same way as Tcl manipulates text strings and GUI elements. This leads to a completely new way of writing the next generation of Web applications. This paper presents the motivation behind the design and implementation of *WebScript* , an overview of its major features, as well as some demonstrations of its power.

## 1 Introduction

*WebScript* is a scripting language for processing Web documents. It is designed and implemented as an extension to Jacl [10, 11], the Java [1] implementation of Tcl [17]. With a concise set of commands and library routines, *WebScript* allows programmers to manipulate HTML [9] in the same way as Tcl manipulates text strings and GUI elements. Meanwhile, the underlying integration of Java and Tcl technologies allows programmers to have all the goodies of both Java and Tcl for free. In addition, *WebScript* provides useful language features like *pipe*, *threads* etc. that greatly simplify the task of programming. We believe that all these features have made *WebScript* an ideal language for quick prototyping and developing the next generation of Web applications.

This paper presents the motivation behind the design and implementation of *WebScript* , an overview of its major language features, as well as some demonstrations of its power.

The rest of the paper is organized as follows: Section 2 presents our motivation behind the design and implementation of *WebScript* . Section 3 overviews the major language features of *WebScript* . Section 4 demonstrates the power of *WebScript* with various practical applications. Section 5 describes related work and compares *WebScript* with alternative approaches for Web document processing. We end with concluding remarks and future work in Section 6.

## 2  Motivation

Since its birth in 1989, the World Wide Web has grown into an enormous popularity. Today it is providing millions of Internet users access to ever-increasing volumes of information. In the mean time, the Web has made the move from an advertising medium to an application platform. Now it supports a tremendous number of Web applications ranging from parcel tracking to online purchasing.

We believe that the next step in Web development is the emergence of composite applications that involve communication among different applications, for instance, a single front end that invokes multiple search engines in parallel and then post-processes the combined results. As the key to the success of building such meta-applications, it is necessary to provide some mechanism for different applications to communicate with each other. This is very difficult due to the enormous number of existing applications.

It is enlightening to look at the success of Internet, because exactly the same problem came up before in the context of Internet and got solved perfectly. Computers on the Internet are interconnected using a wide variety of link technologies ranging from traditional ones like telephone lines, Ethernet, packet radio, and satellite links, to (relatively) modern ones like ATM, FDDI, SMDS and IEEE 802.5 Token Ring. How can these computers communicate with each other? *This is exactly the same problem as the one we've seen before, although the context is different.* The solution to this problem is through the development of the IP protocol. IP became the universal network-level protocol. It unifies all the underlying link technologies and provides a uniform interface. However, it is still too complex for application developers to deal with IP directly. That's why people invented TCP. The central role of TCP is hiding all the complexity of IP and providing a simple picture of network: multiplexed, duplex, connection-oriented, reliable, byte-stream.

TCP/IP's ability to interconnect a variety of networks leads to the explosive growth of the Internet. By July 1998 the Internet had grown to over thirty-six million computers [16]. The tremendous success of Internet indicates a natural solution to our original problem in the context of Web application development: developing a universal interface to unify all the existing applications. Fortunately, we don't have to develop a new one from scratch. We already have an ideal candidate: HTML. Just like IP unifies all underlying link technologies, HTML provides a common interface among various Web applications. Moreover, HTML has already been widely accepted. Almost every Web page we can find today is written in HTML. And many applications like AltaVista etc. are generating

HTML all the time. We believe that HTML will be the universal communication language for the next generation of Web applications, in the same way as IP is the universal network-level protocol. Consequently, having an easy way of manipulating HTML documents will be essential to the success of building Web applications.

However, HTML has its own problems. Various versions and proprietary extensions, loose grammar, large number of syntax errors in existing Web documents ... All these have made it extremely difficult for applications to deal with HTML directly. In the same way as we need TCP over IP, we want to have something that hides the complexity of HTML and provides an easy way of processing HTML content. *WebScript* is designed to address such an intense need. It is designed to allow programmers to manipulate HTML in the same easy way as Tcl manipulates text strings and GUI elements. We believe that this will lead to a completely new way of writing the next generation of Web applications.

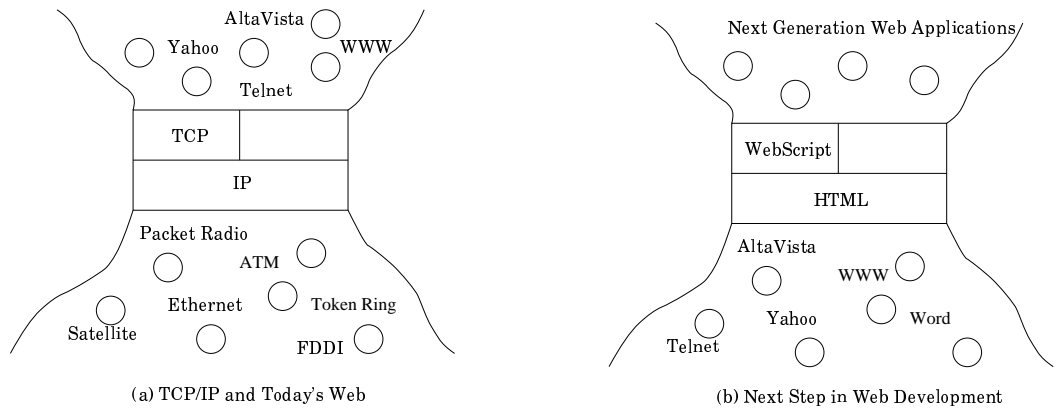

Figure 1. Building Next Generation Web Applications

## 3  *WebScript* – A Scripting Language for the Web

*WebScript* consists of three major components:

- Java Command Language (Jacl)
- Features for processing HTML documents
- Additional useful language features like *pipe* and *thread*.

We overview each of them in the following three subsections.

### 3.1  An Extension to Jacl?

Instead of creating a new language from scratch, we designed and implemented *WebScript* as an extension to Java Command Language, or Jacl, which is a Java implementation of the Tcl scripting language. Our approach has a number of advantages over alternative approaches:

- *WebScript* builds on the syntax and concepts of the Tcl scripting language. Today, with some 500,000 users, Tcl is one of the most popular cross-platform scripting languages. By implementing *WebScript* as an extension to Jacl, we get all the goodies of Tcl for free, for instance, Tcl's extreme extensibility and its superior power for gluing together existing components or applications.

- Implemented in the Java programming language, *WebScript* is extremely portable. It can potentially run on any platform that Java is ported, including Java Station and Palm Pilot. Moreover, as an extension to Jacl, *WebScript* is a Java library that is easy to embed into existing Java applications. With the *WebScript* library, application developers can add HTML capabilities as well as scripting into their Java based applications in a straightforward way.

- Jacl gives programmers direct access to Java objects in a scripting environment. For example, programmers can directly create new instances of any public class and invoke methods on them. This makes Java's powerful GUI features and other built-in libraries directly accessible in Jacl. Jacl also supports the manipulation of JavaBeans, the component mechanism used in the Java platform. *WebScript* inherits all these for free.

## 3.2 Major Features for Web Document Processing

The common tasks in Web document processing can be classified into the following four categories.
- Web document retrieval
- Structure information extraction and representation
- Structured text search and automated navigation
- Page modification

*WebScript* supports all the above by providing a concise set of commands and library routines.

### 3.2.1 Web Document Retrieval

The first step for processing Web documents is to retrieve them from the Web. *WebScript* provides support for Web document retrieval at two levels:

At a lower level, *WebScript* borrows concepts like URL, URL Connection, InputStream, Socket etc. from Java and provides a set of wrapper commands for the corresponding Java classes. These commands allow programmers to provide applications with the same strong network capability as in Java. For example, the following simple piece of *WebScript* code (Figure 2) downloads the home page of Cornell Network Research Group (CNRG) and prints the content:

At a higher level, *WebScript* further simplifies the task of Web development by providing a set of library routines for the common operations in Web document retrieval. Among such higher-level library routines, two most important ones are ws::getpage and ws::postpage. ws::getpage fetches the resource associated with given URL name using HTTP GET method and returns the result as a text string. Users can specify optional query parameters and HTTP header fields, which is especially useful for sending queries to Web search engines. The other routine ws::postpage does almost the same thing except that it uses HTTP POST method instead of HTTP GET. It is mainly used to fill in Web based input forms. The usage of these routines can be best illustrated by the following two examples: Figure 3 does exactly the same thing as Figure 2, but looks much more concise by using ws::getpage. Figure 4 illustrates the usage of optional query parameters.

Most of these higher-level *WebScript* routines are implemented using the lower-level commands we mentioned earlier. This can be well illustrated by the implementation of ws::validate_link (Figure 5), a *WebScript* library routine for link validation:

### 3.2.2 Structure Information Extraction and Representation

The next step is to extract and represent the structure information contained in Web documents.

HTML uses a group of HTML tags to describe the logical structure of a document [9]. HTML tags are simply letter commands surrounded by the $<$ and the $>$ brackets. Within an HTML tag, various *attributes* and *attribute values* can be specified. Most tags in HTML are ended with a similar tag with a slash in it to specify an end to the formatting. Tags are usually nested, forming a mostly hierarchical structure in an HTML document.

The tree data structure is a natural representation for any hierarchical structures. In the past it has been successfully used to organize directories in file systems. For this reason, we choose to represent HTML documents as *tag trees* in *WebScript* . A tag tree consists of a set of interconnected nodes, each of which has one of the three types: text string, comment, or HTML tag. A leaf node can have any type. But any internal node must be an HTML tag, with a tag name and a list of attributes and corresponding attribute values.

The transformation between a tag tree and an HTML document is relatively straightforward if we can parse HTML properly. The tag tree can be constructed at the same time when the document gets parsed. The other direction is even simpler – all we need to do is to make a depth-first traversal on the tree and dump out the content of each node.

All the complexity lies in parsing HTML documents. Most Web documents are not strictly hierarchical due to unbalanced tags and/or syntax errors. Moreover, HTML has a number of different versions and proprietary extensions. All these have made it very difficult to parse HTML documents. We attack these problems with the

```
# create a URL for the home page of CNRG
set u_ [ws::url new "http://www.cs.cornell.edu/cnrg/"]
# create an input stream $ins_ from URL $u_
set ins_ [ws::stream in url $u_]
# read from $ins_ and store the result in string $s_
set s_ [ws::stream read $ins_]
# print string $s_
puts $s_
```

**Figure 2. Getting a page**

```
set s_ [ws::getpage "http://www.cs.cornell.edu/cnrg/"]
puts $s_
```

**Figure 3. Getting a page: a more concise version**

```
# create query parameter list $qp_ with p (query Pattern) = "CNRG" and b (return Base) = "3"
set qp_ [list p "CNRG" b "3"]
# query Yahoo with query parameter list $qp_
set s_ [ws::getpage "http://ink.yahoo.com/bin/query" $qp_]
# print the query result
puts $s_
```

**Figure 4. Query Yahoo for pattern "CNRG" and return 20 matches starting from the $3^{rd}$ match**

```
# validate link with given $urlname
proc ws::validate_link {urlname} {
    set valid_ 0
    catch {
        # create a URL Connection from $urlname
        set c_ [ws::urlconn new $urlname]
        # get status line (0th header field) from $c_
        set statusline_ [ws::urlconn get HeaderField 0 $c_]
        # extract status code
        scan $statusline_ "%s %d" httpver_ statuscode_
        # link valid iff status code is 200. You may use different criteria of course!
        if {$statuscode_ == 200} {
            set valid_ 1
        }
    }
    return $valid_
}
```

**Figure 5. Implementation for ws::validate_link**

following two strategies:

1. Use DTD-driven parser. HTML documents are parsed based on syntax rules specified in the Document Type Definition (DTD). When the HTML version changes, we only need to update the DTD without having to rewrite the whole parser. Notice that DTD-driven parser also allows us to handle Extensible Markup Language (XML) [21] easily, because XML, just like HTML, uses DTD to define its grammar.

2. Be *merciful* for syntax errors. Try to parse as much HTML as possible, rather than breaking down whenever an error occurs.

By hiding all the complexity of HTML, *WebScript* provides programmers with an extremely easy way to transform between HTML and the tag tree representation. Figure 6 illustrates how simple this could be.

---

```
# create a DTD-driven parser $p_
set p_ [ws::parser dtd]
set s_ [ws::getpage http://www.cs.cornell.edu/cnrg/]
# use parser $p_ and DTD "frameset.dtd" to parse string $s_ into a tag tree $tree_
set tree_ [ws::parse $p_ "frameset.dtd" $s_]
# dump $tree_ back into a string with HTML format, but only down to depth 3
set o_ [ws::dump string 3 $tree_]
puts $o_
```

---

**Figure 6. Transformation between HTML and tag tree representation**

Besides parsing and dumping, *WebScript* introduces commands ws::tag, ws::node, ws::parent, ws::child, and ws::sibling to support access to the information in a tag tree. More specifically, ws::tag is used to access and modify the name and attributes in an HTML tag; ws::node allows programmers to create new nodes and modify them; ws::parent, ws::child, and ws::sibling respectively provide access to the parent, child, and sibling nodes of any given node.

### 3.2.3 Structured Text Search and Automated Navigation

We believe that structured text search and automated navigation are two important features for the next generation of Web applications. Today, the common practice for finding documents of interest in the Web is through keyword-based searching and manual navigation (or browsing). Both techniques have well-known limitations. Keyword-based searching completely ignores the structure of the underlying hypermedia and only blindly searches on indexed keywords. This often leads to too many useless hits. Manual navigation is very inefficient and can leads to the infamous *lost-in-hyperspace* syndrome. On the contrary, structured text search, combined with automated navigation, can provide a much better solution by explicitly taking into account the underlying hypermedia structure.

In *WebScript*, automated navigation and structured text search become unified. Since we represent HTML documents as tag trees, structured text search becomes navigation in tag trees by following edges. This is virtually the same as navigation through the Web by following hyperlinks.

To facilitate efficient navigation, *WebScript* supports a set of *structural iterators*, allowing applications to iterate over the nodes of a tag tree in different manners. For example, *WebScript* allows applications to iterate over all nodes in a subtree in either DFS or BFS order. The idea of structural iterator is mainly borrowed from the *iterator* in C++, and is similar to the *Enumeration* in Java, or the *view* in database terminology. Structural iterators make the navigation very simple yet quite efficient. Figure 7 illustrates how we can use structural iterator to extract all hyperlinks in a page.

```
# create a structural iterator iterating in DFS order over all nodes
# with node type "tag" in the subtree rooted at $node_
set it_ [ws::iterator tree dfs "tag" $node_]
# test if $it_ has more elements
while [ws::iterate more $it_] {
    # get next node through $it_
    set n_ [ws::iterate next $it_]
    # get content of $n_ (a tag here)
    set t_ [ws::node get content $n_]
    # test if $t_ is anchor tag (a)
    if {"a" == [ws::tag get name $t_]} {
        # extract hyperlink (the value of attribute "href")
        set link_ [ws::tag get attrib $t_ "href"]
        puts $link_
    }
}
```

**Figure 7. Extract hyperlinks with structural iterator**

### 3.2.4 Page Modification

Many tasks only require modification within each single node. For such simple tasks, basic commands like ws::tag, together with structural iterators, are more than enough. However, there are more complicated tasks involving modification to the tree structure as well. For example, you may want to move a node to a new position.

To support such operations, *WebScript* provides four new commands, namely, ws::cut, ws::copy, ws::paste, and ws::move. Just like what their names indicate, ws::cut cuts a node (subtree) out of a tag tree; ws::copy makes a copy of a node (subtree); ws::paste pastes a copied node (subtree) to some given position; ws::move moves a node (subtree) to a different position without any copying.

With these commands, programmers can manipulate the tree structure in an arbitrary way.

## 3.3 Additional Language Features

*WebScript* provides additional language features to further simplify the task of Web development. Among such features, *pipe* and *thread* are two of the most useful ones.

### 3.3.1 Pipe Command '|'

*WebScript* supports pipe command '|', which is very similar to the standard Unix pipe. Pipe is a method of connecting the output of a command to the input of another. This is a very powerful feature because it allows you to create a very complex command based on basic *WebScript* commands. This makes your programs much more concise. As an example, we can rewrite Figure 4 into the following equivalent but simpler form (Figure 8):

```
| list p "CNRG" b "3"; | ws::getpage "http://ink.yahoo.com/bin/query"; | puts
```

**Figure 8. An equivalent but simpler form for Figure 4**

### 3.3.2 Thread Command ws::thread

Many characteristics of the Web, such as its unreliable services, lack of referential integrity, and its wide area distribution, have made the task of developing Web applications especially difficult. Programmers have to expect and deal with all kinds of situations that do not appear in traditional application development. For example, when we try to retrieve a Web page, it may be unavailable – we've all been tired of the "404 File Not Found" message. Alternatively, it may be redirected to another server and we have to follow the new link to get the page. The server may be unreachable or overloaded and can not handle the request. The connection may be unexpectedly terminated. The data transfer may stall, or may drop to an intolerably low rate.

With *WebScript* command ws::thread, life becomes much easier. First, mechanisms like timeout etc. can be implemented based on ws::thread (as illustrated in Figure 9) to simplify the handling of failures. Moreover, multiple Web services can be accessed in parallel using multiple threads. By exploiting the inherent parallelism of distributed Web servers, we can improve the performance of Web applications.

## 3.4 Implementation

*WebScript* has been implemented entirely in Java as an extension to Jacl. Currently *WebScript* works with JDK versions 1.1 and 1.2 and builds on Solaris, Windows NT, Windows 95, IRIX, Linux, and HPUX.

# 4 Sample Applications in *WebScript*

All the features we described in the previous section have made *WebScript* an ideal language for quick prototyping and developing the next generation of Web applications. In this section, we demonstrate its great power through various practical applications we have built and are currently building with *WebScript* .

## 4.1 *WebGrep*

*WebGrep* , named after the standard Unix command grep, is an application that combines automatic navigation with structured text search. Starting from given URL, *WebGrep* automatically navigates through the underlying hyperlink structure and searches for given pattern on all documents that are reachable within a distance of given number of hyperlinks. For example, we can start from my home page (http://www.cs.cornell.edu/yzhang/) and recursively search for my name "Yin Zhang". The search result is given in Figure 10.

Notice that by specifying appropriate query parameters, we can make the starting URL to be the query result of a Web search engine. Therefore, we can further integrate traditional keyword-based search with structured text search and automatic navigation by using *WebGrep* to post-process the query results returned by traditional search engines.

A simple implementation for *WebGrep* is given in the Appendix. Since we only want to give a flavor of writing applications in *WebScript* , we choose to omit some technical details (for instance, testing whether a URL is associated with an HTML file).

```
# a simple implementation of timeout mechanism
# $timeslot is used for early termination
proc ws::timeout {script timeout {timeslot 500}} {
    # create a thread to run the given $script and start it
    set t_ [ws::thread new]
    ws::thread exec $t_ $script

    # keep looping until either $t_ is done or timeout
    set time_elapsed_ 0
    while {$time_elapsed_ < $timeout} {
        incr time_elapsed_ $timeslot
        # wait for $timeslot milliseconds
        after $timeslot
        set status_ [ws::thread status $t_]
        switch $status_ {
            WS_THREAD_DONE {
                # done => get the result
                set result_ [ws::thread result $t_]
                # garbage collection
                ws::thread destroy $t_
                return $result_
            }
            WS_THREAD_FAIL {
                break
            }
        }
    }
    # timeout => throw an exception
    ws::thread destroy $t_
    error WS_THREAD_FAIL
}
```

**Figure 9. Implementing timeout mechanism with ws::thread**

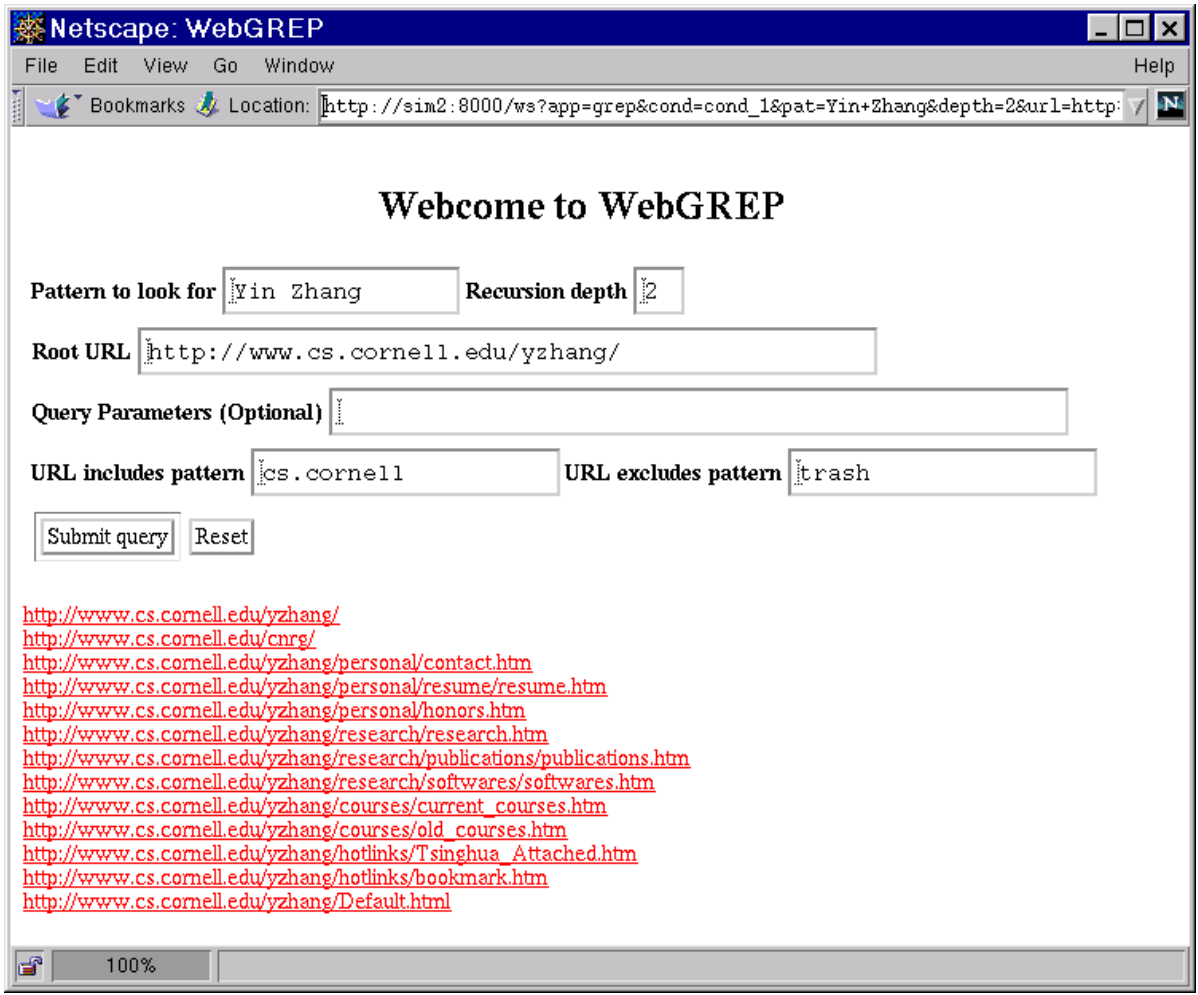

**Figure 10. Sample result returned by** *WebGrep*

## 4.2 Link Validator

We have also implemented a *link validator* in *WebScript* . It can be set as as the proxy for your browser. As you navigate through the Web, the link validator automatically validates all hyperlinks in each page you visit and strikes out all the broken links so that you can easily notice them. Figure 11 illustrates a typical screen snapshot from the link validator. Notice that all the broken links are striken out.

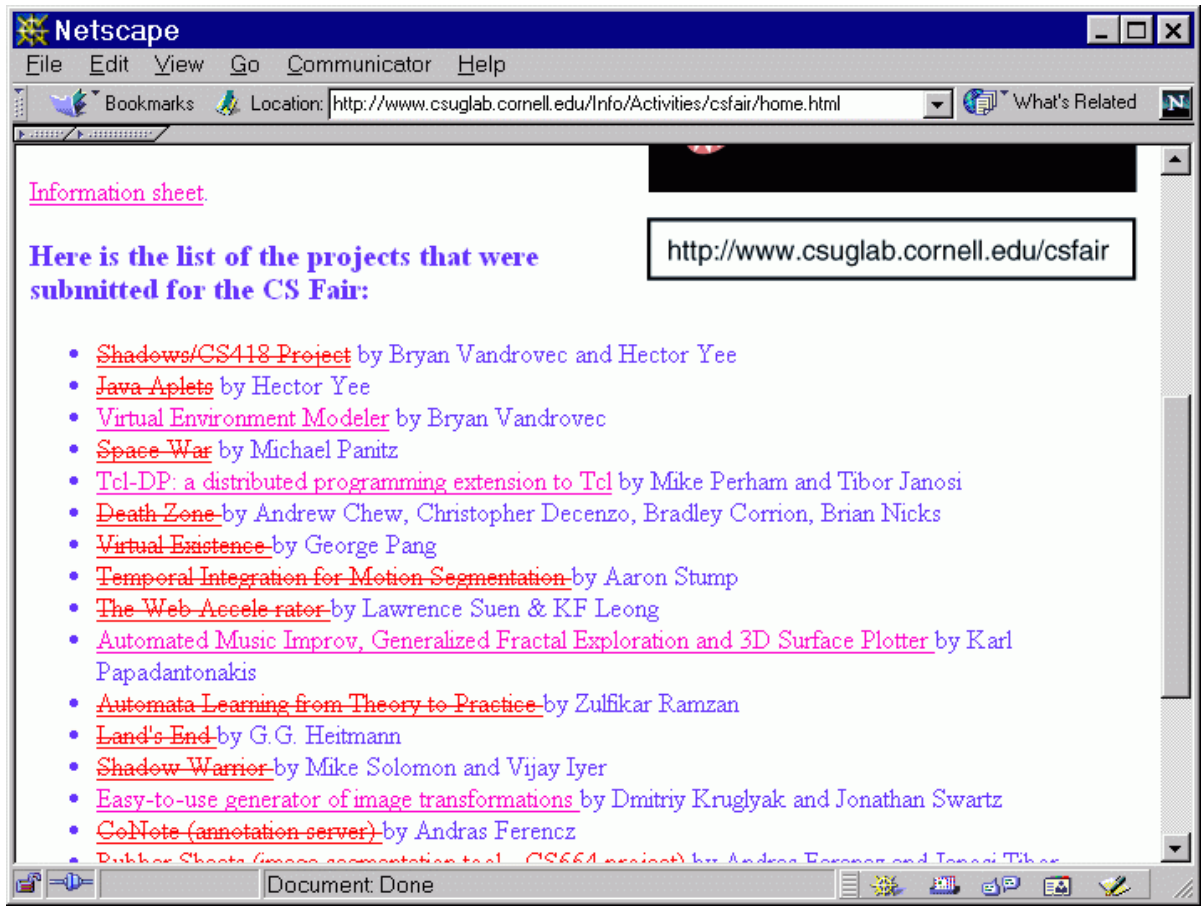

**Figure 11. A Sample Screen Snapshot from the Link Validator**

## 4.3 Other Applications

Other applications we have built and are currently building with *WebScript* include:
- *WebCopy*, a tool that recursively retrieves Web subtrees of HTML documents and related files referenced in HTML anchor tags and stores them locally.
- Meta-Search Engine, a single front-end that invokes multiple search engines on the Web and post-processes the combined result (for instance, find the intersection of the results returned by different search engines.)

- Tools that can "collapse" a page, showing just the headings.
- Tools for automatic creation of site maps and tables of content.
- Tools that allow users to annotate web documents in an arbitrary way. This is similar to what CoNote WWW annotation system [8] does, but with more flexibility.

# 5  Related Work

The HyperText Markup Language (HTML) [9] is the *lingua franca* for publishing hypertext on the World Wide Web. The Extensible Markup Language (XML) [21] is a data format for structured document interchange on the Web. Both HTML and XML use Document Type Definition (DTD) to define grammars and can be handled in a similar way in *WebScript* .

Tool Command Language (Tcl) [17] is one of the most popular scripting languages. Java [1] is one of the most popular system programming languages. Java Command Language (Jacl) [11, 10] is a Java implementation of Tcl. *WebScript* is implemented as an extension to Jacl.

The Standard Document Query Language (SDQL) of the Document Style Semantics and Specification Language [7] introduces the concepts of groves - trees of nodes corresponding to elements in the document. This is similar to our tag tree representation for Web documents.

Web data processing has been actively studied in the context of semi-structured database systems. Stanford's Object Exchange Model (OEM) [18, 3], is a *lightweight object model* developed to act as a general model capable of representing both database and Web data structures. It is first introduced in the TSIMMIS project. A similar model, developed at the University of Pennsylvania, is described in [5, 4]. These models provide a valuable core of ideas for applying database concepts to Web data. Toronto's WebSQL [2] is a declarative query language for extracting information from the Web. The emphasis is to combine query with controlled automated navigation. Technion's WWW Query System (W3QS) [13, 14, 20] views the WWW as an ultra large database and uses W3QL, a high level SQL-like query language, to query the Web. AT&T's UnQL [19] models Web sites as labeled graph and uses UnCAL, a calculus for unstructured data, to express queries for Web sites. Compared with *WebScript* , such database-based approaches are less flexible. They usually require Web data to be extracted into some kind of database. Operations on Web data are limited to database-like operations. Also, as a general programming interface for the Web, they are not as powerful as *WebScript* .

W3C's Document Object Model [6] attempts to specify low-level programming APIs that provide users with the functionality for document navigation and manipulation. However, DOM is restricted to manipulating and searching single HTML and XML elements, does not provide a notion of character patterns, and inherently cannot perform computation.

The Web Interface Definition Language (WIDL) [15] enables automation by mapping Web content into program variables using declarative descriptions of resources. WIDL provides features to submit queries and to extract features from the resulting pages, but page manipulation is not supported.

WebL [12] is also a scripting language for Web document processing. It tries to automate Web development through service combinators and markup algebra. Instead of representing Web documents as trees, WebL uses the data structures of piece and piece set, which in many cases are roughly equivalent to node and node list in a tag tree. Compared to WebL, *WebScript* is much more powerful. WebL's service combinators can be easily implemented as *WebScript* procedures based on the ws::thread command. Moreover, the underlying integration of Tcl and Java in *WebScript* provides goodies of both Tcl and Java for free. This is not achievable in WebL. In addition, due to the lack of explicit tree structures, WebL's operations on Web documents (especially operations for page modification) are relatively "flat".

# 6  Concluding Remarks and Future Work

In this paper we have presented a powerful scripting language called *WebScript* for Web document processing. Implemented as an extension to Jacl, the Java implementation of the Tcl scripting language, *WebScript* provides a number of powerful features that allow programmers to manipulate HTML (XML) in the same way as Tcl manipulates text strings and GUI elements. All these features have made *WebScript* an ideal language for quick prototyping and developing the next generation of Web applications. This can be well demonstrated by the various practical applications that we have built and are currently building.

As our future work, we plan to build more complicated applications with *WebScript* . We also plan to add more useful features into the language.

# 7  Acknowledgements

Special thanks to S. Keshav and Rosen Sharma for their insightful discussion during the initial stage of *WebScript* . Snorri Gylfason suggests the *WebGrep* application. Leo Ku is building applications using *WebScript* .# 8  Appendix: A Simple Implementation for *WebGrep*

---

*# search for $PAT on documents less than $DEPTH hyperlinks away from $URL*

```tcl
proc WebGrep {URL DEPTH PAT} {
    # initialize result to empty
    set result_ [list]
    # $url_list(k) contains all URLs k hyperlinks away from $URL
    set url_list(0) [list $URL]
    # associative array 'seen' records all URLs seen so far
    set seen($URL) 1
    # create a DTD-driven parser using DTD frameset.dtd
    set p_ [ws::parser dtd frameset.dtd]
    for {set i_ 0} {$i_ < $DEPTH} {incr i_} {
        # handle URLs $i_ links away from $URL
        set url_list([expr 1+$i_]) [list]     ; # initialize url_list for next level to empty
        foreach url_ $url_list($i_) {
            # get the page with a timeout of 10000 msec
            if [catch {set page_ [ws::timeout "ws::getpage $url_" 10000]}]{
                continue
            }
            set tree_ [ws::parse $p_ $page_]
            # extract all hyperlinks in $tree_ (see Example 6)
            set it_ [ws::iterator tree dfs "TAG" $tree_]
            while [ws::iterator more $it_] {
                set node_ [ws::iterator next $it_]
                set tag_ [ws::node get content $node_]
                if {[ws::tag get name $tag_] == "A"} {
                    set link_ [ws::tag get attrib $tag_ "href"]
                    if ![info exists seen($link_)] {
                        lappend url_list([expr 1+$i_]) $link_
                        set seen($link_) 1
                    }
                }
            }
            # test if $page_ contains given pattern $PAT
            if [regexp -nocase $PAT $page_] {
                lappend result_ $url_
            }
        }
    }
    return $result_
}
```